\documentclass[epjf,francais]{svjourf}
\usepackage[frenchb]{babel}
\usepackage[utf8]{inputenc}
\usepackage[T1]{fontenc}
\usepackage{lmodern}
\usepackage{epsfig}
\usepackage{amsmath}
\usepackage{amsfonts}
\usepackage{amssymb}
\usepackage{graphicx}
\usepackage{units}
\usepackage{url}
\usepackage[breaklinks=true,colorlinks=true,linkcolor=blue,urlcolor=blue,citecolor=blue]{hyperref}
\usepackage{color}

\newcommand{\RR}{\mathbf{R}}
\newcommand{\rr}{\mathbf{r}}
\newcommand{\xx}{\mathbf{x}}
\newcommand{\kk}{\mathbf{k}}

\newcommand{\dd}{\mathrm{d}}
\newcommand{\ii}{\mathrm{i}}

\newcommand{\ed}{\mathrm{e}}
\newcommand{\rouge}{\color{black}}

\newcommand{\be}{\begin{equation}}
\newcommand{\ee}{\end{equation}}
\newcommand{\bea}{\begin{eqnarray}}
\newcommand{\eea}{\end{eqnarray}}

\DeclareMathOperator\atan{atan}
\DeclareMathOperator\tanhyper{th}
\DeclareMathOperator\im{Im}
\DeclareMathOperator\re{Re}

\begin{document}
\selectlanguage{french}

\title{Mélanges unitaires boson-boson et boson-fermion : troisième coefficient du viriel et paramètre à trois corps sur une résonance de Feshbach étroite} 
\titlerunning{Mélanges unitaires boson-boson et boson-fermion}
\author{Shimpei Endo$^{1,2}$ et Yvan Castin$^{1}$}
\institute{$^{1}$Laboratoire Kastler Brossel, ENS-PSL, CNRS, UPMC-Sorbonne Universit\'es, Coll\`ege de France, Paris, France \\
$^{2}$School of Physics and Astronomy, Monash University, Victoria 3800, Australie}
\authorrunning{Shimpei Endo et Yvan Castin}

\date{6 octobre 2016}

\abstract{Nous donnons des expressions intégrales exactes des troisièmes coefficients d'amas ou du viriel de mélanges binaires de gaz parfaits de bosons ou de fermions, avec {\rouge une} interaction inter-espèce de portée nulle et de longueur de diffusion infinie dans l'onde $s$. En général, le résultat dépend {\rouge de} paramètres à trois corps $R_t$ apparaissant dans des conditions de contact à trois corps, parce qu'un effet Efimov est présent ou parce que le mélange est dans un régime préefimovien avec un rapport de masse proche d'un seuil de l'effet Efimov. Nous donnons une nouvelle expression intégrale exacte de $R_t$ pour le modèle microscopique d'une résonance de Feshbach étroite. Une divergence de $R_t$ dans le régime préefimovien à la valeur $s=1/2$ de l'exposant d'échelle est prédite et physiquement discutée. Les résultats analytiques sont appliqués aux espèces typiques utilisées dans les expériences d'atomes froids.
}

\maketitle

\section{Introduction}
Le domaine des gaz quantiques atomiques a assisté au cours de la dernière décennie à l'émergence d'un nouveau paradigme, le régime des interactions résonnantes, grâce à la possibilité de contrôler à volonté la longueur de diffusion $a$ {\rouge dans l'}onde $s$ {\rouge par} des résonances de Feshbach magnétiques \cite{manips_gen1,manips_gen2,manips_gen3,manips_gen4}. Cela {\rouge  établit un rapport fructueux} entre les corrélations fortes {\rouge à petit nombre de} corps et la physique {\rouge à $N$ corps}, qui n'a guère été étudié auparavant.  À {\rouge l'unitarité}, où $a^{-1}=0$ et l'interaction est invariante d'échelle, il existe encore quelques paramètres {\rouge sur lesquels jouer}, à savoir la statistique quantique et le rapport de masse des particules, qui peuvent affecter profondément les propriétés {\rouge à $N$ corps} du gaz.

Les études expérimentales se sont concentrées jusqu'à présent sur le gaz de {\rouge fermions} de spin 1/2 {\rouge en interaction forte}, et en particulier sur le gaz unitaire de {\rouge fermions}. L'équation d'état du gaz a été mesurée à la fois dans la phase superfluide et dans la phase normale, et les coefficients {\rouge d'amas du développement} de la pression en puissance{\rouge s} de la fugacité, à ne pas confondre avec les coefficients {\rouge du viriel} du {\rouge développement} en puissance{\rouge s} de la densité, ont été extraits jusqu'à l'ordre quatre \cite{Salomon,Zwierlein}. Ces coefficients constituent un {\rouge lien intéressant} entre la physique à {\rouge petit nombre de} corps et celle {\rouge à $N$ corps}.

Bien qu'elles n'aient pas encore été réalisées expérimentalement, des phases moins conventionnelles avec {\rouge toute une variété de} manifestations de la superfluidité ont été proposées, qui tirent parti d'une physique sous-jacente à {\rouge petit nombre de} corps {\rouge très} riche. {\rouge Avec des fermions à plusieurs composantes}, des états trimères ou tétramères peuvent exister \cite{Efimov1,Efimov2,Efimov3,Kartav_discovery} et interagir \cite{EndoNaidontrimer1,EndoNaidontrimer2}, ce qui crée de nouvelles {\rouge voies d'appariement} en concurrence avec l'appariement BCS habituel et {\rouge donne accès} à de nouvelles phases \cite{Nishidatrim1,Nishidatrim2}.  Avec des bosons sans spin en interaction résonnante, des états liés {\rouge à} corps $N$ peuvent exister \cite{vonStecher}, conduisant à une transition liquide-gaz en plus de la transition {\rouge attendue entre un fluide} normal {\rouge et un} superfluide \cite{KrauthPiateckiComparin1,KrauthPiateckiComparin2}. 

Dans cet article, nous présentons un calcul analytique général du troisième coefficient {\rouge d'amas} $b_3$ d'un gaz à deux composant{\rouge e}s à {\rouge l'unitarité} dans l'esprit de \cite{CastinWernerCan} avec la méthode du régulateur harmonique de \cite{ComtetOuvry1,ComtetOuvry2,Drummond1,Drummond2}, en nous appuyant sur le fait que le problème unitaire à trois corps piégés {\rouge peut être résolu} \cite{Jonsell,WernerCastinPRL} {\rouge grâce à} la séparabilité en coordonnées hypersphériques \cite{WernerCastinSym}.  Comme le cas particulier du mélange unitaire fermion-fermion a déjà été traité \cite{ChaoEndoCastin}, nous complétons l'étude en incluant les mélanges unitaires boson-boson et boson-fermion. Nous adhérons ici à la philosophie des modèles {\rouge de} portée nulle, en remplaçant les potentiels d'interaction entre particules par des conditions de contact sur la fonction d'onde à trois corps, à savoir (i) des conditions de contact de Wigner-Bethe-Peierls à deux corps, qui sont invariantes d'échelle à l'unitarité \cite{Wigner,Bethe,CastinCRAS}, et (ii) des conditions de contact à trois corps sur la fonction d'onde {\rouge hyperradiale} : (a) en général, ces conditions sont également invariantes {\rouge d}'échelle (la fonction d'onde hyperradiale {\rouge s'annule} en loi de puissance) ; (b) cependant, si {\rouge un} effet Efimov \cite{Efimov1,Efimov2,Efimov3} {\rouge est présent}, elles doivent {\rouge faire intervenir} une échelle de longueur à trois corps $R_t$, appelée paramètre à trois corps \cite{Efimov1,Efimov2,Efimov3,Danilov}, et une {\rouge coupure} inférieure doit être introduite {\rouge à la main} dans le spectre géométrique des trimères (pour éliminer les états trimères d'énergie de liaison $\approx \hbar^2/m_r R_t^2$ ou plus, où $m_r$ est la masse réduite à deux espèces, car ils sont en général {\rouge hors du régime de validité du} modèle {\rouge de} portée nulle) ;  (c) enfin, si aucun effet Efimov n'est présent, mais que le rapport de masse est suffisamment proche d'un seuil {\rouge de l}'effet Efimov, {\rouge elles} doivent {\rouge tout de même} faire intervenir un paramètre à trois corps $R_t$, de sorte que $b_3$ {\rouge soit une fonction lisse} du rapport de masse à travers le seuil \cite{ChaoEndoCastin} ({\rouge il faut} éliminer également à la main dans ce cas l'état trimère prédit par ces conditions de contact modifiées).

Deux questions importantes doivent cependant être résolues avant de procéder au calcul de $b_3$. La première question est théorique, c'est une lacune du modèle {\rouge de} portée nulle, qui ne peut {\rouge par lui-même} prédire une valeur de $b_3$ lorsqu'un paramètre à trois corps entre en jeu, puisque la connaissance de $R_t$ nécessite un modèle microscopique \cite{EndoNaidonRt1,EndoNaidonRt2,GreeneRt1,GreeneRt2}.  Dans les mélanges boson-boson $\mathrm{B}-\mathrm{b}$, ce problème est aggravé parce que {\sl deux} paramètres à trois corps $R_t$ doivent être introduits, $R_t^{\rm BBb}$ et $R_t^{\rm bbB}$, de sorte que l'on ne peut même pas absorber complètement $R_t$ dans {\rouge un changement} d'échelle de la température comme cela est fait dans \cite{CastinWernerCan}.
  La deuxième question est d'ordre expérimental : lorsque l'effet Efimov est présent dans les expériences actuelles, de fortes pertes à trois corps ont lieu même dans la limite de portée {\rouge nulle} $k_F b\to 0$, où $k_F$ est un nombre d'onde de Fermi et $b$ la portée {\rouge de l}'interaction, en raison de la recombinaison en dimères fortement liés \cite{Petrov_bbo,Petrov_Werner}. Ceci est dû au fait que la probabilité que trois atomes se trouvent dans un {\rouge même volume de} rayon $b$ {\rouge tend vers zéro} trop lentement, seulement comme $b^2$, alors que le taux de recombinaison dans une telle configuration d'atomes proches est $\propto \hbar/m_r b^2$ \cite{estim_pertes}. Jusqu'à présent, cela a limité les études bosoniques {\rouge à $N$ corps} {\rouge à l'}équilibre thermique au régime fortement non dégénéré \cite{SalomonPetrov,Hadzibabic}, où seuls les taux de perte, et non les coefficients {\rouge du viriel}, peuvent être mesurés et comparés à la théorie \cite{SalomonPetrov}, et cela a limité les études bosoniques quantiques dégénérées à un régime {\rouge hors} d'équilibre \cite{Cornell,Chevy,Barth}.

Ici, nous utilisons comme modèle microscopique le modèle d'une résonance de Feshbach (infiniment) étroite \cite{Petrov3be}, où, à {\rouge l'unitarité}, la portée effective de la diffusion à deux corps {\rouge dans l'}onde $s$ est négative ($-2R_*$) et beaucoup plus grande en valeur absolue que la longueur de van der Waals $b$ de l'interaction. Cela résout les deux problèmes susmentionnés. Cela résout d'abord le problème théorique car le problème à trois corps à énergie nulle $E=0^-$ peut être résolu analytiquement et $R_t$ peut être extrait \cite{Mora1,Mora2,Tignone} ; nous affinons la théorie (i) {\rouge en} donnant une expression intégrale de $R_t$ en termes de la fonction {\rouge transcendante} d'Efimov $\bar{\Lambda}_\ell(s)$ (dans \cite{Mora1,Mora2}, une représentation {\rouge en termes d'un} produit infini {\rouge sur les} racines de $\bar{\Lambda}_\ell(s)$ a été donnée ; dans \cite{Tignone}, l'expression intégro-différentielle donnée n'est qu'une {\rouge série} asymptotique), (ii) en étendant la solution d'énergie nulle et le calcul de $R_t$ au cas non {\rouge e}fimovien, pour un rapport de masse proche d'un seuil de l'effet Efimov.  Cela résout également le problème expérimental : le choix d'une résonance de Feshbach étroite dans l'expérience devrait réduire les pertes efimoviennes à trois corps, maintenant dominées par la probabilité d'avoir un atome et une molécule {\rouge de la voie fermée} à une distance $<b$ \cite{LevinsenPetrov}\footnote{La probabilité d'avoir trois atomes {\rouge dans la voie ouverte} à une distance $<b$ avec un moment {\rouge cinétique} relatif $\ell$ est {\rouge un} $O(b^{2\ell+4})$ pour une résonance de Feshbach étroite, voir la {\rouge note} 41 dans \cite{Tignone}.}, où un événement de perte peut avoir lieu avec un taux $\propto \hbar/m_r b^2$ \cite{estim_pertes}. Si l'on considère l'atome et la molécule comme des particules {\rouge discernables} qui n'interagissent pas, comme dans \cite{LevinsenPetrov}, la probabilité de les trouver à {\rouge une} distance $<b$ est {\rouge un} $O(b^3)$ puisque leur fonction d'onde relative est {\rouge un} $O(1)$. On prédit alors un taux de perte à trois corps {\rouge en} $O(k_F b)$, qui {\rouge tend vers zéro} dans la limite de {\rouge portée} nulle. Il reste bien sûr le défi de la stabilisation du champ magnétique {\rouge posé par} l'étroitesse de la résonance.

\section{Troisième coefficient d'{\rouge amas} dans {\rouge d}es modèles {\rouge de} portée nulle}
{\rouge Le développement en amas de} la pression totale $P$ d'un mélange de deux espèces à l'équilibre thermique dans une boîte cubique dans la limite thermodynamique est défini comme {\rouge suit},
\begin{equation}
\frac{P\lambda_r^3}{k_B T} = \sum_{(n_1,n_2)\in\mathbb{N}^{2*}} b_{n_1,n_2} z_1^{n_1} z_2^{n_2}
\end{equation}
 où les fugacités $z_i=\exp(\mu_i/k_B T)$ tendent vers zéro à température {\rouge fixée} $T$, $\mu_i$ est le potentiel chimique de l'espèce $i$ et $\lambda_r=[2\pi\hbar^2/(m_r k_B T)]^{1/2}$ est la longueur d'onde de de Broglie thermique associée à la masse réduite 
\begin{equation}
m_r=\frac{m_1 m_2}{m_1+m_2}
\end{equation}
Dans {\rouge la suite}, nous supposons qu'il n'y a pas d'interaction intra{\rouge -espèce} et que l'interaction inter{\rouge -espèce} est dans la limite unitaire, c'est-à-dire avec une longueur de diffusion infinie {\rouge dans l'onde} $s$ et une portée négligeable. Les coefficients d'amas $b_{n_1,n_2}$ {\rouge du} gaz homogène sont reliés à la limite $B_{n_1,n_2}$ {\rouge lorsque} $\omega\to 0$ des coefficients d'amas du gaz dans des potentiels de piégeage isotropes {\rouge de} même {\rouge pulsation} de piégeage $\omega$ pour les deux espèces \cite{DailyBlume}, 
\begin{equation}
B_{n_1,n_2}=\left(\frac{m_r}{n_1 m_1+n_2 m_2}\right)^{3/2} b_{n_1,n_2}
\label{eq:lienBb}
\end{equation}
{\rouge ce} qui est plus facile à calculer dans la limite unitaire comme {\rouge il est} expliqué dans l'introduction. En raison de l'invariance {\rouge par rotation}, tous les coefficients d'{\rouge amas}, en particulier ceux du troisième ordre, peuvent être écrits comme des sommes sur les contributions des {\rouge différents} secteurs de moment {\rouge cinétique}.  En raison de l'absence d'interaction intra-espèce, $B_{3,0}$ et $B_{0,3}$ ont {\rouge mêmes} valeurs {\rouge que dans un} gaz {\rouge parfait}. Au contraire, $B_{2,1}$ diffère de la valeur {\rouge nulle}, celle d'un gaz {\rouge parfait}. C'est le seul coefficient du troisième ordre non trivial que nous devons calculer puisque $B_{1,2}$ {\rouge s'en déduit} en échangeant le rôle des espèces $1$ et $2$. Nous écrirons sa décomposition sur les secteurs de moment {\rouge cinétique} sous la forme 
\begin{equation}
B_{2,1}=\sum_{\ell\in\mathbb{N}} (2\ell+1) \sigma_\ell
\label{eq:decomp_sur_l}
\end{equation}

Nous nous limiterons {\rouge au cas de} particules {\rouge de faibles} nombres d'onde, beaucoup plus petits que $1/b$, où $b$ est {\rouge désormais la portée} de l'interaction ou {\rouge la portée} effective si {\rouge elle} est plus grande, afin de pouvoir utiliser le modèle à {\rouge portée nulle} de Wigner-Bethe-Peierls, où les interactions entre les espèces sont remplacées par des conditions {\rouge aux} limites sur la fonction d'onde. Le problème unitaire à trois corps qui en résulte peut être résolu analytiquement {\rouge dans} l'espace libre \cite{Efimov1,Efimov2,Efimov3} et dans des pièges harmoniques isotropes \cite{Jonsell,WernerCastinPRL}\footnote{Le problème à trois corps {\rouge dans} l'espace libre peut être résolu analytiquement également pour une longueur de diffusion finie \cite{Macek}.}.  Les fonctions de partition {\rouge canoniques} à trois corps dans le piège et finalement le coefficient $B_{2,1}$ peuvent être calculés. Le résultat s'applique dans le régime de température 
\begin{equation}
k_B T \ll \frac{\hbar^2}{2m_r b^2}
\label{eq:cond1}
\end{equation}

Comme nous allons le voir, un {\rouge ingrédient} central de notre expression analytique est la fonction {\rouge transcendante} d'Efimov. C'est une fonction paire d'une seule variable $s$ qui, dans le secteur {\rouge de} moment {\rouge cinétique} $\ell\in\mathbb{N}$ pour le problème à trois corps $112$, prend la forme\footnote{La fonction {\rouge d'}Efimov est ici divisée par $\cos\nu$ par rapport aux références précédentes, d'où la barre dans la notation $\bar{\Lambda}_\ell(s)$.} \cite{Tignone} : 
\begin{eqnarray}
\label{eq:lampair}
\!\!\!\!\bar{\Lambda}_\ell(s) &\stackrel{\ell\, {\rm pair}}{=}&  1 - \frac{2\eta}{\sin 2\nu}\!\! \int_0^{\nu} \!\! \dd\theta
P_\ell\left(\frac{\sin\theta}{\sin \nu}\right) \frac{\cos (s\theta)}{\cos(s\pi/2)} \\
\!\!\!\!\bar{\Lambda}_\ell(s) &\stackrel{\ell\, {\rm impair}}{=}& 1 + \frac{2\eta}{\sin 2\nu}\!\! \int_0^{\nu}\!\!  \dd\theta
P_\ell\left(\frac{\sin\theta}{\sin \nu}\right) \frac{\sin (s\theta)}{\sin(s\pi/2)}
\label{eq:lamimpair}
\end{eqnarray}
avec $P_\ell$ un polynôme de Legendre, 
\begin{equation}
\nu=\arcsin \frac{m_1}{m_1+m_2}
\end{equation}
l'angle de masse et $\eta=1$ ($\eta=-1$) si l'espèce $1$ est bosonique (fermionique). Une écriture équivalente et {\rouge qui a son utilité} peut être obtenue à partir de la forme hypergéométrique de \cite{GasaneoMacekBirse1,GasaneoMacekBirse2} : 
\begin{multline}
\bar{\Lambda}_\ell(s)=1-\frac{\eta(-\sin\nu)^\ell}{2\pi^{1/2}\cos\nu} \\ 
\times\sum_{k\in\mathbb{N}}\frac{\Gamma(k+\frac{\ell+1+s}{2})\Gamma(k+\frac{\ell+1-s}{2})}
{\Gamma(k+\ell+\frac{3}{2})} \frac{\sin^{2k}\nu}{k!}
\label{eq:forme2}
\end{multline}
où $\Gamma$ est la fonction Gamma d'Euler.

Comme nous le verrons, dans le domaine des modèles {\rouge de} portée {\rouge nulle}, le modèle de Wigner-Bethe-Peierls n'est pas {\rouge l'alpha et l'oméga}, car il ne spécifie que des conditions de contact à deux corps. Pour choisir le modèle {\rouge de} portée {\rouge nulle} approprié pour le problème à trois corps, il faut discuter de l'existence d'une racine de $\bar{\Lambda}_\ell(s)$ sur l'intervalle $]0,\ell+1[$. Sur cet intervalle, $\bar{\Lambda}_\ell(s)$ est une fonction lisse de $s$ car le plus petit pôle positif des termes de la série (\ref{eq:forme2}) {\rouge se trouve en} $\ell+1$. {\rouge Pour mener cette} discussion, on peut profiter d'un premier résultat utile : la fonction $s\mapsto \bar{\Lambda}_\ell(s)$ est décroissante (croissante) sur $]0,\ell+1[$ lorsque $\eta (-1)^\ell$ est positif (négatif)\footnote{{\rouge Pour le voir, on peut} utiliser l'éq.~(\ref{eq:forme2}) et le fait que {\rouge la fonction} $x\mapsto \Gamma(x+a)\Gamma(a-x)$ est positi{\rouge ve} et a une dérivée logarithmique {\rouge strictement positive} $\sum_{n\in\mathbb{N}} \frac{2x}{(a+n)^2-x^2}$ sur $[0,a[$ $\forall a>0$, voir \S 8.362(1) dans \cite{GR}.} et tend vers $-\infty$ ($+\infty$) lorsque $s\to \ell+1^-$. Par conséquent, elle {\rouge admet zéro ou} une racine dans $]0,\ell+1[$, selon le signe de $\bar{\Lambda}_\ell(0)$.  Nous {\rouge disposons} d'un deuxième résultat utile : la fonction $S\mapsto \bar{\Lambda}_\ell(\ii S)$ est croissante (décroissante) sur $\mathbb{R}^+$ lorsque $\eta (-1)^\ell$ est positi{\rouge f} (négati{\rouge f})\footnote{{\rouge En effet,} $\forall a>0$, la dérivée logarithmique de $x\mapsto \Gamma(a+\ii x)\Gamma(a-\ii x)$ sur $\mathbb{R}$ est $\sum_{n\in\mathbb{N}} \frac{-2x}{(n+a)^2+x^2}$, voir \S 8.362(1) dans \cite{GR}.} et tend exponentiellement rapidement vers $1$ {\rouge en} $+\infty$ {\rouge en vertu d}es éqs.~(\ref{eq:lampair},\ref{eq:lamimpair}). On n'est alors confronté qu'à l'un des trois cas possibles énumérés ci-dessous ; pour $\eta (-1)^\ell$ négatif, seul le premier cas est effectivement accessible puisque $\bar{\Lambda}_\ell(s)$ est alors $>1$ sur $]0,\ell+1[$.

{\sl 1. Le cas simple non efimovien} : $\bar{\Lambda}_\ell(s)$ n'a que des racines réelles, et la plus petite racine positive est $s_\ell >1$. La fonction d'onde hyperradiale $F(R)$ est soumise à la condition {\rouge aux} limites\footnote{Plus précisément, la fonction d'onde à trois corps s'écrit sous la forme $\psi(\rr_1,\rr_2,\rr_3)=\phi(\Omega)F(R)/R^2$ où $\Omega$ est l'ensemble des hyperangles \cite{Efimov1,Efimov2,Efimov3,WernerCastinSym}.} 
\begin{equation}
F(R)\underset{R\to 0}{=} O(R^{s_\ell})
\label{eq:condnonefim}
\end{equation}
où l'hyper{\rouge rayon} $R$ du système $112$ est l'{\rouge écart type} (pondéré par {\rouge les} masses correspondantes) des positions des trois particules {\rouge comptées à partir de} leur centre de masse. Alors la contribution $\sigma_\ell$ {\rouge de} moment {\rouge cinétique} $\ell$ à $B_{2,1}$ {\rouge comme la} défini{\rouge t} l'éq.~(\ref{eq:decomp_sur_l}) est donnée par \cite{CastinWernerCan,ChaoEndoCastin} 
\begin{equation}
\sigma_\ell=-\int_{\mathbb{R}^+}\frac{\dd S}{2\pi} \ln \bar{\Lambda}_\ell(\ii S)
\end{equation}

{\sl 2. Le cas pr{\rouge é}efimovien} : $\bar{\Lambda}_\ell(s)$ n'a que des racines réelles, mais la plus petite racine positive $s_\ell\in ]0,1[$.  Dans la {\rouge voie} associée à la racine $s_\ell$, il existe une condition {\rouge aux} limites enrichie \cite{FelixThese,Nishidareson,Safavi}, qui doit être utilisée lorsque $s_\ell$ est suffisamment petit \cite{ChaoEndoCastin}\footnote{On ne l'utilisera pas pour $s_\ell>1$ car la fonction d'onde n'est pas {\rouge de carré intégrable} {\rouge en} $R=0$, $\int_0^{R_0} \dd R\, R |F(R)|^2=+\infty$ pour tout $R_0$ fini.} : 
\begin{equation}
F(R)\underset{R\to 0}{=} (R/R_t)^{s_\ell} - (R/R_t)^{-s_\ell} + O(R^{2-s_\ell})
\label{eq:condpreefim}
\end{equation}
où la longueur $R_t$ est un paramètre à trois corps. {\rouge Ceci conduit à prédire} dans l'espace libre un état lié $112$ de dégénérescence $2\ell+1$ et d'énergie de liaison 
\begin{equation}
\label{eq:eglob_preefim}
E_{\rm glob} = \frac{2\hbar^2}{(2m_1+m_2)R_t^2}
\left(\frac{\Gamma(1+s_\ell)}{\Gamma(1-s_\ell)}\right)^{1/s_\ell} 
\end{equation}
qui doit être {\rouge ignoré} en l'absence de résonance à trois corps. Alors \cite{ChaoEndoCastin} 
\begin{equation}
\sigma_\ell=-\int_{\mathbb{R}^+}\frac{\dd S}{2\pi} f_\ell(S)
-\int_{\mathbb{R}^+} \dd\epsilon \Delta(\epsilon) \beta \ed^{-\beta\epsilon}
\end{equation}
avec $\beta=1/k_B T$, 
\begin{equation}
\Delta(\epsilon) = \frac{1}{\pi} \atan \frac{\tanhyper [\frac{s_\ell}{2} \ln (\epsilon/E_{\rm glob})]}{\tan (\frac{s_\ell}{2}\pi)}
\label{eq:Delta_revis}
\end{equation}
et la fonction lisse à valeur{\rouge s} réelle{\rouge s} sur l'axe réel : 
\begin{equation}
f_\ell(S)\equiv\ln\left[\frac{S^2+1}{S^2+s_\ell^2}\bar{\Lambda}_\ell(iS)\right]
\label{eq:def_f}
\end{equation}
Dans ce cas pré{\rouge e}fimovien, le modèle {\rouge de} portée nulle {\rouge enrichi} a la même condition de validité que l'éq.~(\ref{eq:cond1}) sous l'hypothèse générique que $R_t$ et $b$ sont du même ordre de grandeur \cite{ChaoEndoCastin}.

{\sl 3. le cas {\rouge e}fimovien} : $\bar{\Lambda}_\ell(s)$ a des racines réelles et une paire de racines {\rouge imaginaires pures} $\pm s_\ell$ avec $s_\ell = \ii |s_\ell|$. Dans la voie {\rouge e}fimovienne, on doit utiliser la condition aux limites \cite{Efimov1,Efimov2,Efimov3,Danilov} 
\begin{equation}
F(R) \underset{R\to 0}{=} (R/R_t)^{\ii |s_\ell|} - (R/R_t)^{-\ii |s_\ell|}+O(R^2)
\label{eq:condefim}
\end{equation}
Comme $F(R)$ a un nombre infini de {\rouge racines sur tout voisinage de} $R{\rouge =}0$, le modèle prédit dans l'espace libre un nombre infini d'états trimères d'énergies $\epsilon_q$ et de dégénérescence $2\ell+1$, formant une {\rouge suite} géométrique que nous tronquons à la main pour la rendre cohérente avec l'hypothèse d'{\rouge une interaction de} portée nulle : 
\begin{equation}
\epsilon_q=-E_{\rm glob} \ed^{-2\pi(1+q)/|s_\ell|}, \forall q\in\mathbb{N}
\label{eq:epsq}
\end{equation}
L'échelle d'énergie globale est un {\rouge simple prolongement} analytique de l'éq.~(\ref{eq:eglob_preefim}) : 
\begin{equation}
\label{eq:eglob_efim}
E_{\rm glob}= \frac{2\hbar^2}{(2m_1+m_2)R_t^2} \ed^{[\ln \Gamma(1+s_\ell)-\ln \Gamma(1-s_\ell)]/s_\ell}
\end{equation}
avec $\ln\Gamma$ la {\rouge détermination} habituelle du logarithme de la fonction $\Gamma$. En plus de la condition de portée {\rouge nulle} (\ref{eq:cond1}), on exige pour que le modèle {\rouge considéré} soit {\rouge valable} que le trimère {\rouge fondamental} soit dans le régime de portée {\rouge nulle}, 
\begin{equation}
|\epsilon_0|\ll E_{\rm glob}
\end{equation}
c'est-à-dire $|s_\ell|$ $\lesssim 1,\!5$. Comme $|s_\ell|$ est une fonction croissante du rapport de masse $m_1/m_2$, {\rouge ceci impose une limite} supérieure {\rouge sur} $m_1/m_2$. Alors \cite{ChaoEndoCastin} 
\begin{eqnarray}
\sigma_\ell &=& -\int_{\mathbb{R}^+}\frac{\dd S}{2\pi} f_\ell(S) -\int_{\mathbb{R}^+} \dd\epsilon \Delta(\epsilon) \beta \ed^{-\beta\epsilon} \nonumber \\
&+& \sum_{q\in\mathbb{N}} \left(\ed^{-\beta\epsilon_q}-1\right)
\label{eq:sigmal_cas3}
\end{eqnarray}
où {\rouge la fonction} $f_\ell(S)$ est toujours donné{\rouge e} par l'éq.~(\ref{eq:def_f}) et la nouvelle forme de la fonction $\Delta$, \footnote{\label{note:autre_forme}
On {\rouge dispose également de l'expression} $\Delta(\epsilon)=\frac{|s_\ell|x}{2\pi}+\frac{1}{\pi}\im \ln \left(1-\ed^{-\pi |s_\ell|}\ed^{-\ii |s_\ell|x}\right)$ \cite{relgenbos}, qui conduit à $-\int_{\mathbb{R}^+} \dd\epsilon \Delta(\epsilon)\beta\ed^{-\beta\epsilon}=
\frac{|s_\ell|}{\pi}\{\frac{1}{2}\ln(\ed^\gamma \beta E_{\rm glob})-\sum_{n\in\mathbb{N}^*} \ed^{-n\pi|s_\ell|}
\re[\Gamma(-\ii n |s_\ell|) (\beta E_{\rm glob})^{\ii n |s_\ell|}]\}$ comme dans \cite{CastinWernerCan}, $\gamma=0,\!577\, 215\ldots$ étant la constante d'Euler.} 
\begin{equation}
\Delta(\epsilon)= \frac{1}{\pi} \atan \frac{\tan (\frac{|s_\ell|}{2}x )}{\tanhyper (\frac{|s_\ell|}{2}\pi)}
+\left\lfloor \frac{|s_\ell|x}{2\pi}\right\rceil
\label{eq:Delta_efim}
\end{equation}
reste une fonction lisse de $x=\ln(\epsilon/E_{\rm glob})$ grâce à la fonction {\rouge entier} le plus proche {\rouge introduite} dans le dernier terme.

\begin{figure}[tbp]
\begin{center}
\includegraphics[width=0.8\linewidth,clip=]{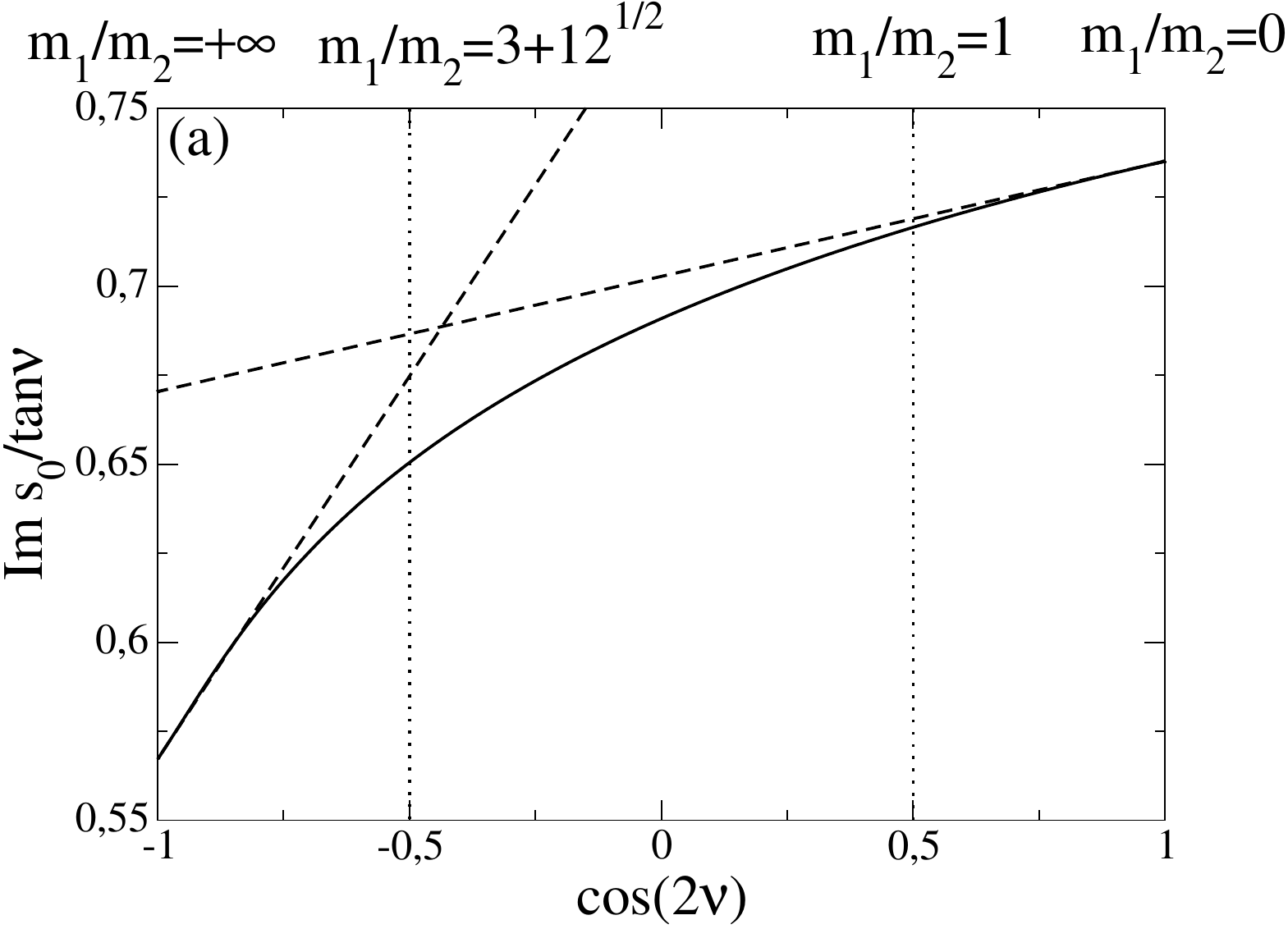}  \includegraphics[width=0.7\linewidth,clip=]{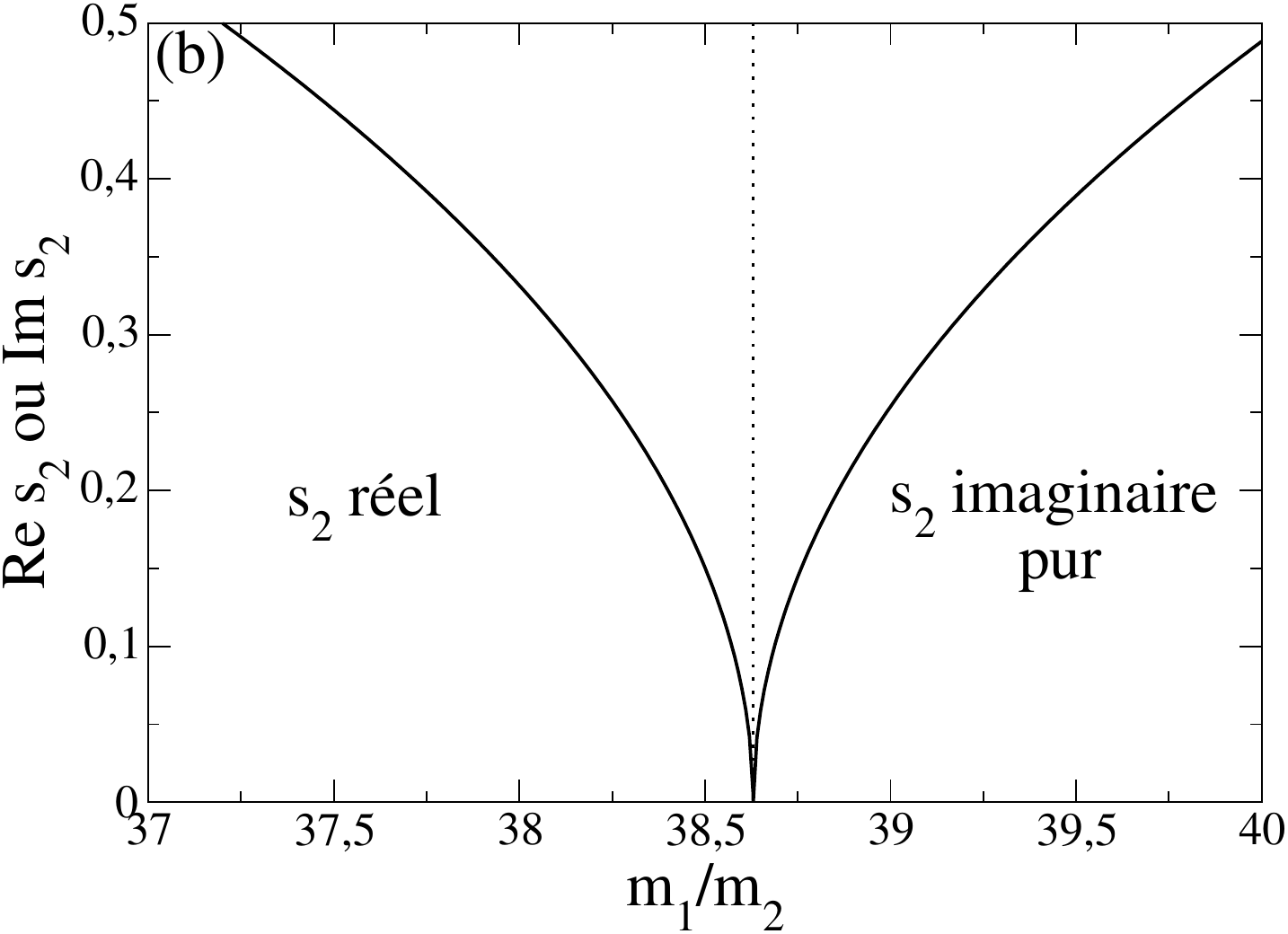}  \end{center}
\caption{Partie réelle ou imaginaire de $s_\ell$ en fonction du rapport de masse $m_1/m_2$ dans le problème $112$ pour $\eta=1$ (les particules de l'espèce $1$ sont des bosons identiques) dans le secteur {\rouge de} moment {\rouge cinétique} (a) $\ell=0$ ($s_0$ est toujours imaginaire {\rouge pur}), (b) $\ell=2$ ($s_2$ passe de réel à imaginaire {\rouge pur} pour $m_1/m_2=38,\!6301\ldots$ {\rouge sur} la ligne pointillée verticale).  Dans (a) on utilise l'angle de masse $\nu=\arcsin\frac{m_1}{m_1+m_2}\in [0,\frac{\pi}{2}]$ comme paramètre, pour {\rouge pouvoir} couvrir tous les rapports de masse, et $\im s_0$ est divisé par $\tan\nu$ pour former une quantité bornée. {\rouge Traits} pleins : solution numérique de l'équation {\rouge transcendante} $\bar{\Lambda}_\ell(s_\ell)=0$. Lignes {\rouge tiretées} en (a) : développements {\rouge limités} pour un faible rapport de masse $\nu\to 0^+$, $\im s_0/\tan\nu=\frac{4}{\pi\sqrt{3}}\left[1+\left(\frac{1}{3\pi^2}-\frac{7}{90}\right)(1-\cos 2\nu)+O(\nu^4)\right]$, et pour un grand rapport de masse $\nu\to \frac{\pi}{2}^-$ comme dans \cite{Tignone}, $\im s_0/\tan\nu=\Omega\left[1+\frac{3+\Omega}{6(1+\Omega)}(1+\cos 2\nu)\right]+O[(\frac{\pi}{2}-\nu)^4]$ où la constante $\Omega$ {\rouge satisfait} à $\Omega \exp\Omega=1$. Lignes pointillées verticales en (a) : les rapports de masse $m_1/m_2=1$ et $3+\sqrt{12}$ où $\cos 2\nu =\pm \frac{1}{2}$.} 
\label{fig:sl}
\end{figure}

Les conditions sur le rapport de masse pour {\rouge qu'il y ait} un effet Efimov sont connues.  Pour $\eta=+1$, il y a un effet Efimov dans le secteur $\ell=0$ pour {\rouge tout} rapport de masse, dans le secteur $\ell=2$ pour $m_1/m_2>38,\!6301\ldots$, dans le secteur $\ell=4$ pour $m_1/m_2>125,\!764\ldots$, etc. Pour $\eta=-1$, il y a un effet Efimov dans le secteur $\ell=1$ pour $m_1/m_2>13,\!6069\ldots$ \cite{Petrov2003}, dans le secteur $\ell=3$ pour $m_1/m_2>75,\!9944\ldots$ \cite{Kartavtsev}, etc. {\rouge Sur} la fig.~\ref{fig:sl}, {\rouge faite pour} $\eta=1$, nous {\rouge représentons} la partie imaginaire de $s_0$ en fonction de l'angle de masse (avec une comparaison à ses développements {\rouge limités} à petit et grand rapport de masse) ; nous {\rouge représentons} également les parties réelle et imaginaire de $s_2$ près du seuil de l'effet Efimov.

\section{Paramètre à trois corps pour une résonance de Feshbach étroite}

Pour le modèle d'interaction microscopique {\rouge d'une} résonance de Feshbach étroite, nous montrons maintenant comment obtenir une nouvelle expression analytique pour le paramètre à trois corps et nous {\rouge précisons} l'intervalle de rapport de masse {\rouge sur} lequel {\rouge il faut} utiliser le modèle pré{\rouge e}fimovien enrichi {\rouge plutôt que le} modèle non efimovien simple.

Le point de départ est l'équivalent de l'équation intégrale de Skorniakov--Ter-Mar\-ti\-ro\-sienne \cite{TerMartirosian} pour le problème à trois corps $112$ sur une résonance de Feshbach étroite. Elle peut être obtenue en utilisant {\rouge soit} des conditions de contact à deux corps modifiées \cite{Petrov3be}, soit un modèle à deux {\rouge voies} avec des potentiels séparables de portée tendant vers zéro \cite{Tignone,Burnett,WernerCastinmol,PricoupenkoLasinio,Zwerger,Pricoupenkoseul}.  Dans le secteur {\rouge de} moment {\rouge cinétique} $\ell$, les deux modèles conduisent pour $1/a=0$ et une énergie négative $E=-\hbar^2q^2/2m_r$ dans le {\rouge référentiel} du centre de masse à 
\begin{multline}
[q_{\rm rel}(k)+q_{\rm rel}^2(k)R_*] d(k)=\\
\eta \int_{\mathbb{R}^+} \frac{\dd k'}{\pi} \int_{-1}^{1} \dd u \frac{P_\ell(u) k'^2 d(k')}{q^2+k^2+k'^2+2kk'u\sin\nu}
\label{eq:TerM}
\end{multline}
avec $P_\ell$ le polynôme de Legendre de degré $\ell$, $q_{\rm rel}(k)\equiv|q+\ii k\cos\nu|$ et $R_*>0$ la longueur de Feshbach {\rouge entrant dans} l'amplitude de diffusion $12$ à l'énergie $E$ dans le {\rouge référentiel} du centre de masse, $f_q=-1/(\ii q+q^2 R_*)$. On a {\rouge tenu compte} de l'invariance {\rouge par rotation} avec l'ansatz suivant {\rouge sur} la fonction inconnue : 
\begin{equation}
D(\kk)=Y_\ell^m(\hat{\kk}) d(k)
\end{equation}
où $\hat{\kk}\equiv \kk/k$ est la direction de $\kk$ et $Y_\ell^m$ est une harmonique sphérique. Dans le modèle à deux {\rouge voies} (en supposant pour des raisons de simplicité l'absence d'interaction directe {\rouge dans la voie ouverte}), $D(\kk)$ est l'amplitude de probabilité {\rouge dans la voie fermée} d'avoir une particule de vecteur d'onde $\kk$ et une molécule de vecteur d'onde $-\kk$. Dans le modèle de \cite{Petrov3be} $D(\kk)$ est la transformée de Fourier de la partie régulière $A(\xx)$ de la fonction d'onde, 
\begin{equation}
D(\kk)=\int\dd^3 x\, \ed^{-\ii \kk\cdot\xx} A(\xx)
\label{eq:DTFA}
\end{equation}
telle que $\psi(\rr_1,\rr_2,\rr_3)\sim
\frac{A(\rr_3-\RR_{12})}{r_{12}}$ lorsque la distance im\-pu\-re\-té-première particule identique $r_{12}\to 0$ à des positions fix{\rouge é}es $\RR_{12}$ de leur centre de masse et $\rr_3$ de la seconde particule identique. Dans la limite de {\rouge portée nulle}  et dans le secteur {\rouge de} moment {\rouge cinétique} $\ell$, elle est reliée à la fonction d'onde hyperradiale $F(R)$ {\rouge comme suit,} 
\begin{equation}
A(\xx)\propto Y_\ell^{m}(\hat{\xx}) x^{-1} F\left(R=\frac{x\sin^{1/2}\nu}{1+\sin\nu}\right)
\label{eq:AefdF}
\end{equation}

Dans la limite {\rouge d'}énergie {\rouge nulle} $E=0^-$, il y a invariance d'échelle de l'opérateur intégral dans l'éq.~(\ref{eq:TerM}), qui devient un produit de convolution lorsqu'on utilise $X=\ln(kR_*\cos\nu)$ comme nouvelle variable. On prend alors la transformée de Fourier de $d(k)k^2$ par rapport à $X$ comme nouvelle fonction inconnue ; elle obéit à une équation {\rouge aux} différence{\rouge s} finie{\rouge s} {\rouge que l'on sait résoudre}. En {\rouge prenant} la transformée de Fourier {\rouge inverse} et en utilisant le théorème des résidus, on peut calculer analytiquement le comportement de $d(k)$ {\rouge pour} $X\to-\infty$, c'est-à-dire {\rouge pour} $k\to 0$, ce qui permet d'accéder au paramètre à trois corps $R_t$.

Cette procédure a déjà été mise en œuvre dans \cite{Mora1,Mora2,Tignone} dans le cas efimovien $s_\ell=\ii |s_\ell|$. Nous {\rouge la} poussons jusqu'à l'ordre {\rouge sous-dominant} pour obtenir une condition de validité {\rouge du} modèle {\rouge de} portée nulle : 
\begin{multline}
\!\!\!\! d(k)\underset{k\to 0}{=} C(-s_\ell) (kR_*\cos\nu)^{-s_\ell-2}\!\! \left[1-\frac{kR_*\cos\nu}{\bar{\Lambda}_\ell(1-s_\ell)}\right]+  \\
C(s_\ell) (kR_*\cos\nu)^{s_\ell-2}\!\! \left[1-\frac{kR_*\cos\nu}{\bar{\Lambda}_\ell(1+s_\ell)}\right]+o(kR_*)^{-1}
\label{eq:dkdev}
\end{multline}
avec\footnote{Une meilleure estimation du reste dans l'éq.~(\ref{eq:dkdev}) est $O(kR_*)^\alpha$ avec $\alpha=\min(u_{\ell,1}-2,0)$ pour $s_\ell\in\ii\mathbb{R}^+$ et $\alpha=-s_\ell$ pour $s_\ell\in ]0,1[$ (auquel cas $\ell\geq 1$).} 
\begin{multline}
C(s)=\Gamma(-2s) \frac{\Gamma(1+v_{\ell,0}+s)}{\Gamma(v_{\ell,0}-s)}  \\
\times \prod_{n\in\mathbb{N}^*} \frac{\Gamma(u_{\ell,n}-s)\Gamma(1+v_{\ell,n}+s)}
{\Gamma(v_{\ell,n}-s)\Gamma(1+u_{\ell,n}+s)}
\label{eq:defC}
\end{multline}
où les $u_{\ell,n}$, $n\geq 1$, sont les racines positives et $v_{\ell,n}=\ell+1+2n$, $n\geq 0$, les pôles positifs de $\bar{\Lambda}_\ell(s)$ {\rouge rangés dans l'}ordre croissant. En utilisant les éq{\rouge s}.~(\ref{eq:condefim},\ref{eq:AefdF}) pour obtenir $A(\xx)$ dans le modèle {\rouge de} portée nulle et en {\rouge en prenant la transformée de} Fourier comme dans l'éq.~(\ref{eq:DTFA}), on obtient\footnote{On utilise $\int_0^{+\infty} \dd x\, x^{\frac{1}{2}} J_{\ell+\frac{1}{2}}(x) x^{-s}=
2^{\frac{1}{2}-s}\frac{\Gamma(\frac{2+\ell-s}{2})}{\Gamma(\frac{1+\ell+s}{2})}$, où la fonction de Bessel $J$ provient du développement d'une onde plane {\rouge sur les} harmoniques sphériques.} 
\begin{equation}
\left(\frac{q_{\rm glob} R_*}{2}\right)^{2s_\ell}=\frac{C(-s_\ell)}{C(s_\ell)} \prod_{k=0}^{\ell} \frac{k-s_\ell}{k+s_\ell}
\label{eq:qglobprod}
\end{equation}
où 
\begin{equation}
E_{\rm glob}\equiv \frac{\hbar^2 q_{\rm glob}^2}{2m_r}
\label{eq:Eglob_en_fonction_de_qglob}
\end{equation}
est {\rouge re}lié au paramètre à trois corps par l'éq.~(\ref{eq:eglob_efim}). Ceci {\rouge généralise} la référence \cite{Tignone} au cas de deux bosons et d'une impureté. {\rouge De façon} remarquable, nous avons trouvé une nouvelle forme analytique ne nécessitant pas le calcul des racines $u_{\ell,n}$, $n\geq 1$, de la fonction {\rouge transcendante} $\bar{\Lambda}_\ell(s)$:\footnote{On utilise l'identité $\frac{\Gamma(u+s)\Gamma(u+1+s)\Gamma(v-s)\Gamma(v+1-s)}{\Gamma(u-s)\Gamma(u+1-s)\Gamma(v+s)\Gamma(v+1+s)}=
\exp\left\{-\ii \int_{\mathbb{R}^+} \frac{\dd x}{\tanhyper\pi x} [\ln\Psi(x-\ii s)-\ln\Psi(x+\ii s)]\right\}$ avec $\Psi(z)=\frac{u^2+z^2}{v^2+z^2}$ et $u,v>0 $. Elle est valable pour {\rouge tout} $s\in\ii\mathbb{R}^+$, et aussi pour $0< s<u,v$. {\rouge Pour l'établir, on remarque qu'}elle est clairement {\rouge satisfaite en} $s=0$, et {\rouge que} sa dérivée logarithmique par rapport à $s$ {\rouge s'obtient à partir} de la formule \S 8.361(3) dans \cite{GR}, $\frac{\dd}{\dd z} \ln \Gamma(z)=\ln z-\frac{1}{2z} -\int_{\mathbb{R}^+} \frac{2t \dd t}{(t^2+z^2)[\exp(2\pi t)-1]}$ avec $\re z>0$.} 
\begin{multline}
\left(\frac{q_{\rm glob} R_*}{2}\right)^{2s_\ell}=\frac{\Gamma(1+2s_\ell)\Gamma(v_{\ell,0}-s_\ell)\Gamma(2-s_\ell)}
{\Gamma(1-2s_\ell)\Gamma(v_{\ell,0}+s_\ell)\Gamma(2+s_\ell)} \\
\times \exp\left[-\ii\int_0^{+\infty} \dd S\, \frac{f_\ell(S-\ii s_\ell)-f_\ell(S+\ii s_\ell)}{\tanhyper(\pi S)}\right]
\label{eq:qglobana}
\end{multline}
où la fonction $f_\ell$ est donnée par l'éq.~(\ref{eq:def_f}). Ceci est aussi explicite que possible puisque $\bar{\Lambda}_\ell(s)$ {\rouge peut être écrit} explicitement grâce aux éq{\rouge s}.~(\ref{eq:lampair},\ref{eq:lamimpair})\footnote{Pour $s_\ell\in \ii \mathbb{R}^+$, on prend le logarithme de l'éq.~(\ref{eq:qglobana}) pour exprimer $q_{\rm glob}R_*$ de manière univoque en termes de la {\rouge détermination} habituelle de la fonction $\ln \Gamma(z)$, et on peut {\rouge utiliser} la forme numériquement plus commode $2\ii \int_0^{|s_\ell|} \dd S f_\ell(S)+\ii \int_{\mathbb{R}^+} \dd S \left(1-\frac{1}{\tanhyper(\pi S)}\right) [f_\ell(S+|s_\ell|)-f_\ell(S-|s_\ell|)]$ pour l'argument de la fonction exponentielle dans {\rouge l'é}q.~(\ref{eq:qglobana}).}.

La {\rouge méthode de résolution} de l'équation intégrale à énergie nulle peut être étendue au régime préefimovien $0<s_\ell<1$.\footnote{Le seul point subtil est {\rouge qu'il faut} déplacer la ligne d'intégration définissant la représentation de Fourier de $k^2 d(k)$ considérée comme une fonction de $X$ : au lieu d'intégrer sur la ligne $\mathbb{R}+\ii 0^+$ comme dans l'éq.~(66) de \cite{Tignone}, on doit intégrer sur une ligne $\mathbb{R}+\ii\gamma$, $s_\ell < \gamma < 1+v_{\ell,0}$, pour passer entre le pôle le plus élevé $\ii s_\ell$ de la transformée de Fourier provenant d'une racine de $\bar{\Lambda}_\ell$ et le pôle le plus bas $\ii (1+v_{\ell,0})$ provenant d'un pôle de $\bar{\Lambda}_\ell$.} Tous les résultats (\ref{eq:dkdev}-\ref{eq:qglobana}) du cas {\rouge e}fimovien s'étendent directement au cas préefimovien.  Pour $0<s_\ell<1/2$, le {\rouge second membre} de l'éq.~(\ref{eq:qglobana}) est positif. Pour $1/2<s_\ell<1$, il est négatif ; cela signifie que $R_t^{2s_\ell}$ est négatif, ce qui conduit dans l'éq.~(\ref{eq:condpreefim}) à une condition {\rouge aux} limites $F(R)\underset{R\to 0}{=}(R/|R_t|)^{s_\ell}+(R/|R_t|)^{-s_\ell}+O(R^{2-s_\ell})$ ne {\rouge conduisant} plus {\rouge à} un état lié à trois corps comme discuté dans \cite{FelixThese}. 

Il est intéressant de noter qu'à un rapport de masse $m_1/m_2$ tel que $s_\ell=\frac{1}{2}$, $q_{\rm glob}$ et $1/R_t$ {\rouge s'annulent} comme $1-2s_\ell$ en raison du premier facteur $\Gamma$ {\rouge au} dénominateur de l'éq.~(\ref{eq:qglobana}) et à {\rouge la propriété} $\Gamma(z)\sim 1/z$ pour $z\to 0$, {\rouge ce qui} suggère une résonance à trois corps. S'il s'agissait d'une véritable résonance à trois corps, cependant, on croirait à la condition {\rouge aux} limites (\ref{eq:condpreefim}) à toute échelle d'énergie beaucoup plus petite que l'échelle d'énergie $\hbar^2/(2m_r R_*^2)$ {\rouge associée à la portée de l}'interaction ; pour $s_\ell-\frac{1}{2}$ petit et négati{\rouge f}, il y aurait {\rouge alors} un état trimère d'énergie de liaison $E_{\rm glob}$. {\rouge Or,} ceci est en contradiction avec une {\rouge ré}solution numérique de l'éq.~(\ref{eq:TerM}) pour $\eta=-1$ \cite{Tignone} et pour $\eta=1$.  Pour identifier la {\rouge bonne} échelle d'énergie $E_{\rm lim}$ en dessous de laquelle les prédictions du modèle {\rouge de} portée nulle {\rouge enrichi} (\ref{eq:condpreefim}) {\rouge sont} fiables, nous exigeons que pour $k_{\rm lim}\cos\nu \equiv (2m_r E_{\rm lim})^{1/2}/\hbar$ le terme {\rouge sous-dominant} de l'éq.~(\ref{eq:dkdev}) soit comparable au terme {\rouge dominant}\footnote{La condition correspondante pour $s_\ell\leftrightarrow -s_\ell$ {\rouge met en jeu} $\bar{\Lambda}_\ell(1+s_{\ell})$ et est moins {\rouge contraignante}.} : 
\begin{equation}
k_{\rm lim} R_*\cos\nu = |\bar{\Lambda}_\ell(1-s_\ell)|
\end{equation}
Comme $\bar{\Lambda}_\ell(s)\propto s-s_\ell$ {\rouge près} de sa racine $s_\ell$, $\bar{\Lambda}_\ell(1-s_\ell)$ {\rouge s'annule} comme $1-2s_\ell$ quand $s_\ell\to 1/2$, donc $k_{\rm lim}\cos\nu \simeq q_{\rm glob}$ et $E_{\rm lim}\simeq E_{\rm glob}$.  Ceci explique l'absence d'état trimère même pour $q_{\rm glob}R_*\ll 1$.  On peut donc calculer le coefficient d'amas de $112$ sur une étroite résonance de Feshbach {\rouge près} de $s_\ell=1/2$ en utilisant le modèle {\rouge de} portée {\rouge nulle} {\rouge enrichi} (\ref{eq:condpreefim}) seulement sous la condition 
\begin{equation}
k_B T \ll E_{\rm glob}
\label{eq:cond2}
\end{equation}
Lorsque $s_\ell$ est très proche de $1/2$, $k_B T$ devient en pratique $>E_{\rm glob}$ et on doit revenir au modèle habituel {\rouge de} portée {\rouge nulle} (\ref{eq:condnonefim}), sous la condition de validité (\ref{eq:cond1}) écrite pour $b=R_*$. L'équation~(\ref{eq:dkdev}) devient alors 
\begin{multline}
d(k)\underset{k\to 0}{\stackrel{s_\ell=\frac{1}{2}}{=}} (kR_*\cos\nu)^{-5/2} - (k R_*\cos\nu)^{-3/2} [\mbox{cte} \\ +\ln(kR_*\cos\nu)]
+O[\ln(kR_*\cos\nu) (k R_*\cos\nu)^{-1/2}]
\end{multline}
Le terme {\rouge dominant} correspond à la condition {\rouge aux} limites habituelle (\ref{eq:condnonefim}), et le terme {\rouge sous-dominant} est {\rouge effectivement} négligeable {\rouge (à un facteur logarithmique près)} {\rouge sous la condition} $k R_*\cos\nu \ll 1$.

\begin{figure}[tbp]
\begin{center}
\includegraphics[width=0.8\linewidth,clip=]{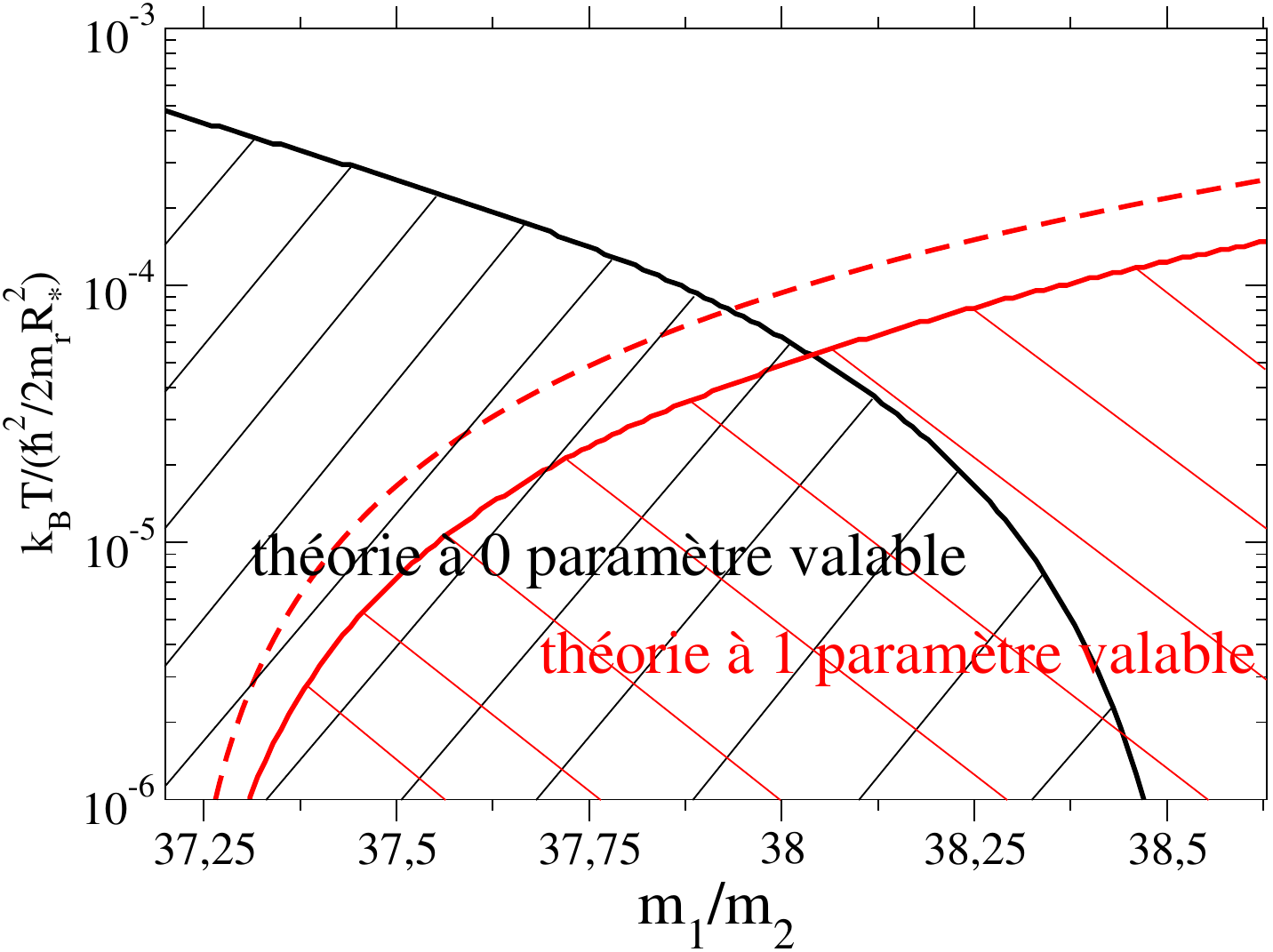}  \end{center}
\caption{Pour $\eta=+1$ dans le secteur {$\ell=2$}, zone de validité dans le plan (rapport de masse, température) de la théorie {\rouge de portée nulle} habituelle {\rouge à zéro paramètre} (hachur{\rouge es} noires) et de la théorie {\rouge de portée nulle} enrichie {\rouge à un paramètre} (hachur{\rouge es} rouges) pour le calcul du coefficient d'{\rouge amas} $B_{2,1}$ à l'unita{\rouge rité} sur une résonance de Feshbach étroite de longueur {\rouge de Feshbach} $R_*$. Sous la ligne noire {\rouge en trait plein}, la somme des termes de degré $-s_2-1$ et $s_2-2$ {\rouge en} $k$ est plus de 3 fois plus petite que le terme de degré $-s_2-2$ dans l'éq.~(\ref{eq:dkdev}), pour $k \cos\nu=(2m_r k_B T)^{1/2}/\hbar$. Sous le trait plein rouge, la somme des termes de degré $-s_2-1$ et $s_2-1$ dans $k$ est plus de 3 fois plus petite que la somme des termes de degré $-s_2-2$ et $s_2-2$. Ligne pointillée rouge : {\rouge là où} $k_B T=E_{\rm glob}$. Aux {\rouge bornes} de l'intervalle de masse {\rouge considéré}, {\rouge on a} respectivement $s_2=1/2$ et $s_2=0$.}
\label{fig:regimval}
\end{figure}

Pour $s_\ell$ significativement éloigné de $1/2$ par {\rouge valeurs inférieures}, ou pour $s_\ell$ imaginaire {\rouge pur}, les modèles {\rouge de} portée {\rouge nulle} {\rouge enrichis} (\ref{eq:condpreefim},\ref{eq:condefim}) prédisent toujours l'existence d'un état trimère d'énergie de liaison $E_{\rm glob}$, ce qui est incorrect sur une résonance de Feshbach étroite. Cela signifie que ces modèles ne sont jamais corrects à l'échelle d'énergie $E_{\rm glob}$.  Dans le calcul {\rouge des coefficients du} viriel, nous prenons donc comme condition de validité qualitative 
\begin{equation}
k_B T \ll  E_{\rm glob}, \frac{\hbar^2}{2m_r R_*^2}
\label{eq:condvalquali}
\end{equation}
{\rouge ce} qui inclut automatiquement la contrainte obtenue ci-dessus au voisinage de $s_\ell=1/2$. Les régimes de température {\sl quantitativement} déterminés où le modèle habituel {\rouge de portée nulle} à zéro paramètre et le modèle {\rouge de portée nulle} enrichi à {\rouge un} paramètre peuvent être utilisés pour calculer le troisième coefficient d'amas pour un rapport de masse conduisant à $s_\ell\in [0,1/2]$ sont représentés {\rouge sur} la fig.~\ref{fig:regimval}, en prenant comme exemple le cas bosonique $\eta=+1$ dans le secteur $\ell=2$. Ils confirment l'analyse {\sl qualitative} ci-dessus.

\begin{figure}[tbp]
\begin{center}
\includegraphics[width=0.74\linewidth,clip=]{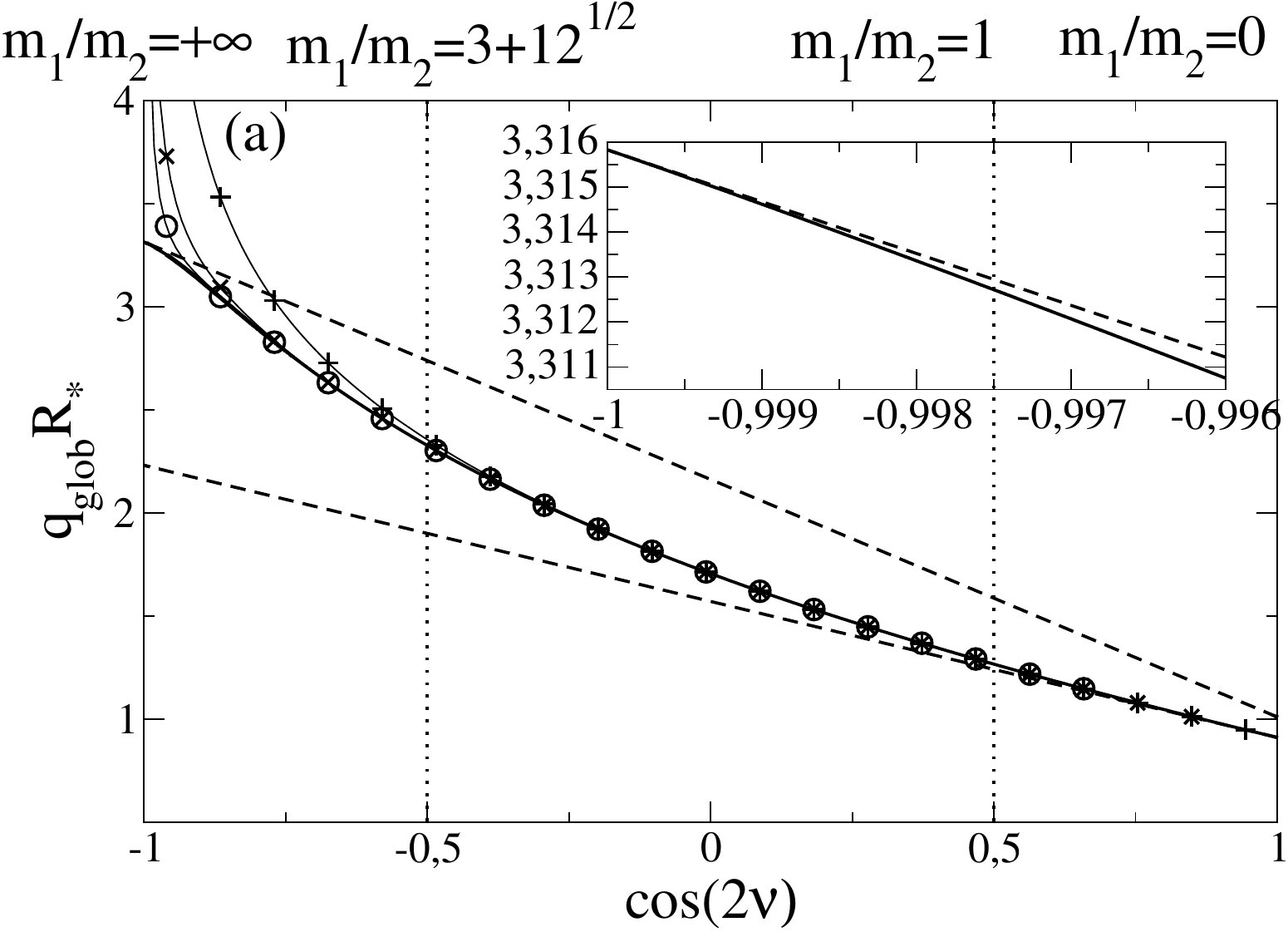}  \includegraphics[width=0.74\linewidth,clip=]{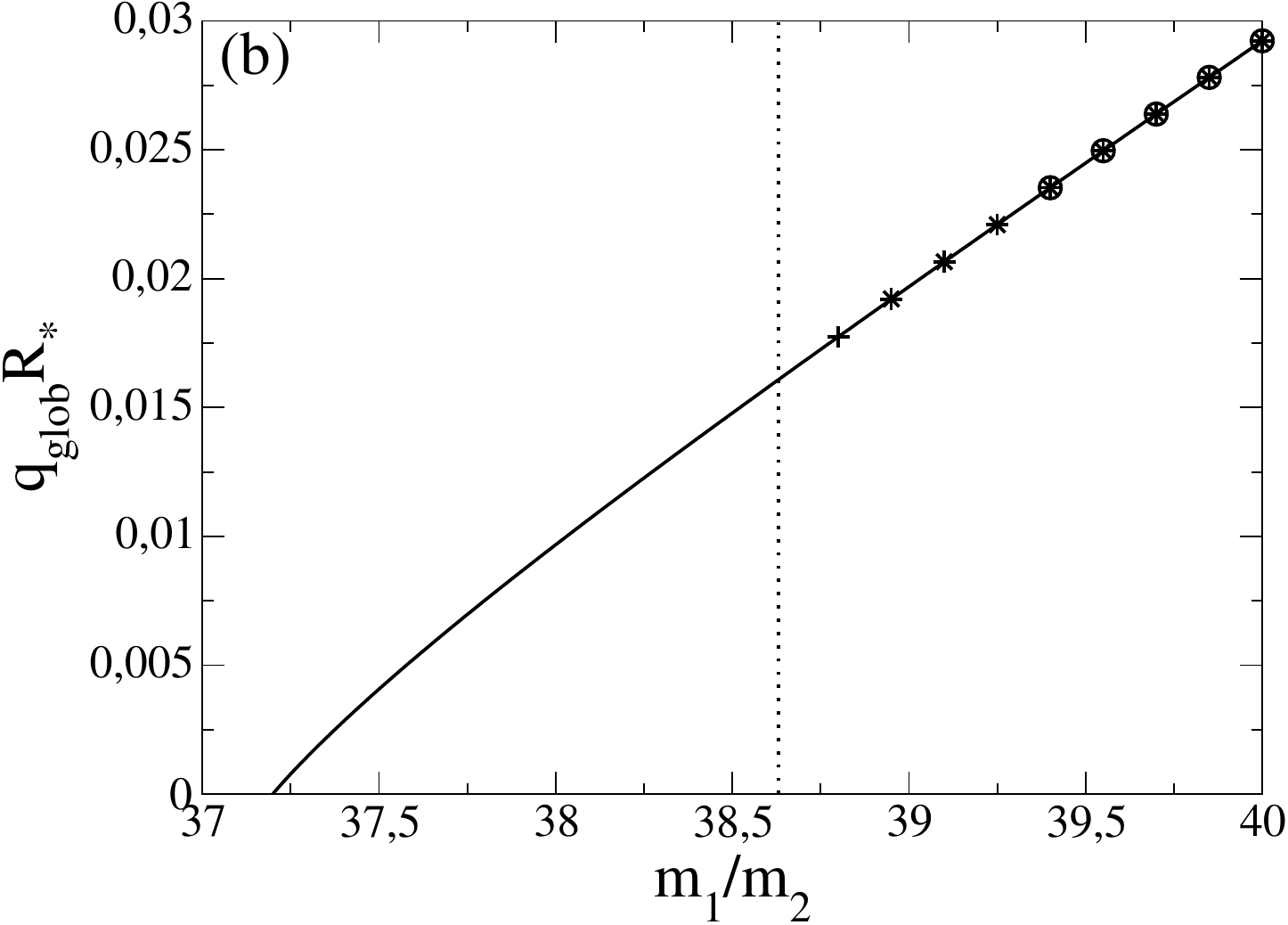}  \end{center}
\caption{Valeur de $q_{\rm glob}R_*$ donnant l'échelle d'énergie globale $E_{\rm glob}=\hbar^2 q_{\rm glob}^2/2m_r$ [ou de manière équivalente le paramètre à trois corps $R_t$ selon les éqs.~(\ref{eq:eglob_preefim},\ref{eq:eglob_efim})] pour deux bosons identiques {\rouge de l'}espèce $1$ et une particule d'espèce différente $2$ interagissant avec une longueur de diffusion infinie {\rouge dans} l'onde $s$ sur une résonance de Feshbach étroite, en fonction du rapport de masse $m_1/m_2$ ou de l'angle de masse $\nu=\arcsin \frac{m_1}{m_1+m_2}$, dans le secteur {\rouge de} moment {\rouge cinétique} $\ell=0$ pour (a) et $\ell=2$ pour (b). {\rouge Trait plein} épais : valeur exacte {\rouge é}q.~(\ref{eq:qglobana}), à la fois pour le cas pr{\rouge é}efimovien limité à $s_\ell\in ]0,1/2]$ [dans (b) à gauche de la ligne pointillée verticale] et pour le cas efimovien $s_\ell\in \ii \mathbb{R}^+$ [partout dans (a) et à droite de la ligne pointillée verticale dans (b)].  Symboles {\rouge reliés par des traits} fins : dans le cas {\rouge e}fimovien, à partir des estimations $E_{\rm glob}\simeq -\epsilon_q \exp[2\pi (q+1)/|s_\ell|]$ (exactes pour $q\to +\infty$ ou $|s_\ell|\to 0$), où $\epsilon_q$ est l'énergie du $q$ème état trimère, $q\in\mathbb{N}$, obtenue numériquement à partir de l'éq.~(\ref{eq:TerM}) quand c'est {\rouge faisable} ($q=0$ : plus, $q=1$ : croix, $q=2$ : cercles) ; quand $m_1/m_2\to+\infty$ la partie basse du spectre devient hydrogénoïde \cite{Tignone} (plutôt que géométrique) et $(-2m_r \epsilon_q)^{1/2}R_*/\hbar\to +\infty$, d'où les {\rouge écarts} à la {\rouge courbe en trait plein noir épais}. En (a) les droites obliques en {\rouge tiretés} sont les {\rouge développements} (\ref{eq:devqglob0}) et (\ref{eq:dev2qglob0}), ce dernier étant en réalité utile sur une plage très limitée, voir {\rouge l'agrandissement} dans l'encadré, et les lignes pointillées verticales {\rouge repèrent les} rapports de masse $m_1/m_2=1$ et $3+\sqrt{12}$ où $\cos 2\nu =\pm \frac{1}{2}$. En (b), $q_{\rm glob}$ {\rouge s'annule} au rapport de masse tel que $s_2=1/2$, mais il ne s'agit pas d'une véritable résonance à trois corps (voir le texte). La longueur de Feshbach $R_*$ est liée à la portée effective à deux corps par $r_e=-2R_*$ à l'unitarité. \label{fig:qglob}}
\end{figure}

{\rouge Sur} la fig.~\ref{fig:qglob}, nous avons représenté le paramètre à trois corps $R_t$, ou plus commodément $q_{\rm glob}R_*$, en fonction du rapport de masse $m_1/m_2$, dans le cas bosonique $\eta=1$, à tous les angles de masse {\rouge pour $\ell=0$} et au voisinage du seuil de l'effet Efimov pour $\ell=2$.  Du côté efimovien, nous comparons également $E_{\rm glob}$ aux {\rouge quantités} $-\epsilon_q \exp[2\pi (q+1)/|s_\ell|]$, pour les premiers états trimères $q=0$, $q=1$, etc. 
du modèle de résonance étroite de Feshbach, qui {\rouge dans la} précision de la théorie {\rouge de portée nulle} coïncident avec $E_{\rm glob}$.

Avec notre nouvelle expression (\ref{eq:qglobana}), certains résultats analytiques explicites pour le paramètre à trois corps peu\-vent être obtenus dans la limite d'une {\rouge impureté (particule de l'espèce 2)} lourde $m_2/m_1\gg 1$ ou légère $m_2/m_1\ll 1$, {\rouge en considérant} le cas {\rouge où les} particules {\rouge de l'espèce} $1$ {\rouge sont} bosoniques ($\eta=+1$) {\rouge et en se limitant au} secteur $\ell=0$. Dans la limite d'{\rouge une} impureté lourde $\nu\to 0^+$, nous obtenons le développement 
\begin{multline}
\frac{q_{\rm glob}R_*}{2} \underset{\nu\to 0^+}{=} \ed^{-\pi/4}\left[1+ \left(\frac{17}{27}- \frac{1}{9\pi^2} + \frac{19\pi}{72}\right) \sin^2\nu\right.\\
\left. \phantom{\left(\frac{0}{27}\right)} +O(\sin^4\nu) \right]
\label{eq:devqglob0}
\end{multline}
Dans la limite d'{\rouge une} impureté légère $\nu\to \frac{\pi}{2}^-$, le développement analytique prend la forme\footnote{Dans l'intégrale sur $S$ sous l'exponentielle de l'éq.~(\ref{eq:qglobana}), on doit traiter séparément la contribution de $S=O(1)$ et de $S\approx |s_0|$. La première contribution conduit dans $\ln(q_{\rm glob}R_*)$ à un terme linéaire en $\pi/2-\nu$ qui {\rouge compense} exactement celui provenant du logarithme des facteurs $\Gamma$ dans le préfacteur de l'éq.~(\ref{eq:qglobana}).} : 
\begin{multline}
\label{eq:dev2qglob0}
\frac{q_{\rm glob}R_*}{2}\underset{\nu\to\frac{\pi}{2}^-}{=} \ed^{J_0}\left\{1+\left[\frac{2\Omega J_2}{3(\Omega+1)} +\frac{1}{12\Omega^2}\right.\right. \\  
\left.\left. -\frac{2}{3\Omega}+\frac{1}{24(1+\Omega)}\right] \cos^2\nu +O(\cos^3\nu)\right\}
\end{multline}
où nous avons introduit $\Omega=0,\!567\, 143\ldots$ {\rouge solution de} $\Omega \exp\Omega=1$ et les intégrales 
\begin{eqnarray}
J_0 &\equiv& \int_0^1 \dd\tau \ln \frac{\ed^{\Omega \tau}+\tau-1}{\tau} = 0,\!505\, 560\ldots \\
J_2 &\equiv& \int_0^1 \dd\tau \frac{\tau \ed^{\Omega\tau}}{\ed^{\Omega\tau}+\tau-1} = 0,\!194\, 862\ldots
\end{eqnarray}
Le terme {\rouge dominant} {\rouge au second membre} de l'éq.~(\ref{eq:dev2qglob0}) est en accord avec \cite{Tignone}, le terme {\rouge sous-dominant} est nouveau. Les {\rouge développements pour} $\nu\to 0^+$ et $\nu\to \frac{\pi}{2}^-$ sont représentés par des lignes {\rouge tiretées} {\rouge sur} la fig.~\ref{fig:qglob}a.

\section{Application à des mélanges atomiques {\rouge concrets}}
\begin{table*}
\label{table:quantbb}
\caption{Pour toutes les combinaisons possibles des espèces bosoniques ${}^{7}$Li, ${}^{41}$K et ${}^{87}$Rb, les quantités physiques du problème à trois corps $112$ qui {\rouge entrent dans} le troisième coefficient d'{\rouge amas} {\rouge correspondant} $B_{2,1}$ (voir le texte). Le cas $\mathrm{X}-\mathrm{X}'$, avec un rapport de masse unit{\rouge é}, correspond à un isotope atomique donné pris dans deux états internes différents. L'interaction entre les espèces $1$ et $2$ est décrite par le modèle de résonance Feshbach étroite de longueur Feshbach $R_*$. L'énergie correspondante du trimère de l'état fondamental $\epsilon_0$ [lorsqu'elle n'est pas trop {\rouge faible} pour être obtenue numériquement à partir de l'éq.~(\ref{eq:TerM})] et l'échelle d'énergie globale $E_{\rm glob}$ liée au paramètre à trois corps par l'éq.~(\ref{eq:eglob_efim}) sont données en unités de $\hbar^2/2m_r R_*^2$. Pour les rapports de masse considérés, l'effet Efimov {\rouge est présent} dans le secteur $\ell=0$ seulement, et {\rouge on a} $s_\ell>1$ dans tous les autres secteurs. La partie non efimovienne $B_{2,1}^{\rm non\, efim}$ de $B_{2,1}$ est la somme du premier terme de l'éq.~(\ref{eq:sigmal_cas3}) pour $\ell=0$ et de toutes les contributions de $\ell>0$ dans l'éq.~(\ref{eq:decomp_sur_l}), elle est indépendante de la température et {\rouge du} paramètre {\rouge à} trois corps.}
\begin{center}
\begin{tabular}{cccccccc}
\hline
espèce 1 & espèce 2 & $m_1/m_2$ & $\im s_0$ & $-\epsilon_0$ & $-\epsilon_0 e^{2\pi/|s_0|}$ & $E_{\rm glob}$  & $B_{2,1}^{\rm non\, efim}$\\
\hline
${}^7$Li & ${}^{87}$Rb & $0,\!080728$ & $0,\!055037$ & -- & -- & $0,\!845022$  & $-0,\!13645$ \\
${}^7$Li & ${}^{41}$K & $0,\!17128$ & $0,\!108458$ & $6,\!12\times 10^{-26}$ & $0,\!884(1)$ & $0,\!884068$ & $-0,\!14692$ \\
${}^{41}$K & ${}^{87}$Rb & $0,\!47132$ & $0,\!246214$ & $9,\!13\times 10^{-12}$ & $1,\!104669$ & $1,\!104669$ & $-0,\!17285$\\
$\mathrm{X}$ & $\mathrm{X}'$ & 1 & $0,\!413697$ &  $4,\!07 \times 10^{-7}$ & $1,\!606453$ & $1,\!606449$ & $-0,\!20539$ \\
${}^{87}$Rb & ${}^{41}$K & $2,\!12171$ & $0,\!644404$ & $1,\!55\times 10^{-4}$ & $2,\!663 601$ & $2,\!662428$ & $-0,\!25361$ \\
${}^{41}$K & ${}^7$Li & $5,\!8383$ & $1,\!073851$ & $1,\!50\times 10^{-2}$ & $5,\!221479$ & $5,\!125277$ & $-0,\!31954$\\
${}^{87}$Rb & ${}^7$Li & $12,\!3873$ & $1,\!521051$ & $1,\!29\times 10^{-1}$ & $8,\!038473$ & $7,\!329207$ & $-0,\!20990$ 
\end{tabular}
\end{center}
\end{table*}

\begin{figure}[htbp]
\begin{center}
\includegraphics[width=0.95\linewidth,clip=]{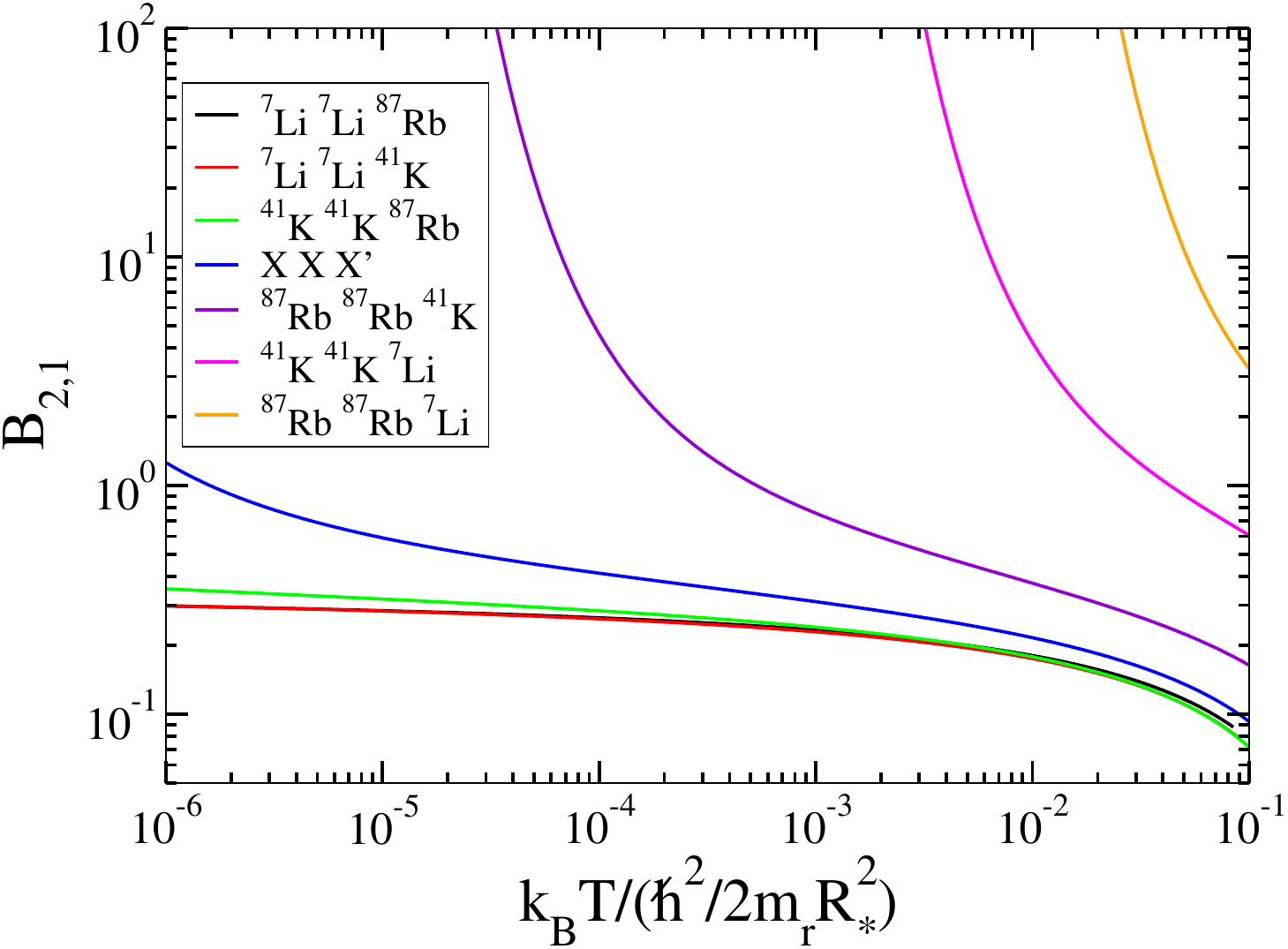}  \end{center}
\caption{{\rouge Troisième} coefficient d'amas {\rouge dans le régulateur} harmonique $B_{2,1}$ en fonction de la température, pour des mélanges binaires unitaires d'espèces atomiques bosoniques ${}^{7}$Li, ${}^{41}$K et ${}^{87}$Rb {\rouge couramment utilisées} dans {\rouge les} expériences {\rouge sur les} gaz froids. Il n'y a pas d'interaction intra{\rouge -espèce}. L'interaction inter-espèce est traitée dans le modèle {\rouge de} portée {\rouge nulle} avec des conditions de contact à deux et trois corps invariantes {\rouge d}'échelle, sauf dans le secteur efimovien $\ell=0$ où les conditions à trois corps {\rouge font intervenir} un paramètre $R_t$ comme dans l'éq.~(\ref{eq:condefim}), {\rouge dont la} valeur {\rouge est} obtenue à partir du modèle de résonance de Feshbach étroite de longueur de Feshbach $R_*$, voir les éq{\rouge s}.~(\ref{eq:eglob_efim},\ref{eq:qglobana}). Le modèle {\rouge de} portée nulle a une applicabilité limitée au régime de basse température (\ref{eq:condvalquali}), et à des rapports de masse atomique pas trop grands de sorte que le trimère {\rouge fondamental} reste dans la limite de portée nulle. Ici, $m_r=\frac{m_1m_2}{m_1+m_2}$ est la masse réduite des deux espèces, et XXX$'$ correspond à un mélange de deux états internes différents du même isotope atomique.}
\label{fig:concretebb}
\end{figure}

Pour illustrer les expressions analytiques obtenues dans cet article, nous {\rouge représentons} {\rouge sur} la fig.~\ref{fig:concretebb} le {\rouge troisième} coefficient d'{\rouge amas} {\rouge dans le régulateur} harmonique $B_{2,1}$ en fonction de la température pour des mélanges boson-boson ($\eta=1$) unitaires réalistes, en {\rouge considérant} les espèces atomiques ${}^{7}$Li, ${}^{41}$K et ${}^{87}$Rb qui ont déjà été refroidies expérimentalement à {\rouge très basse} température.  Nous incluons le cas d'un rapport de masse unit{\rouge é} car il correspond à la même espèce atomique prise dans deux états internes différents. Nous rappelons que les coefficients d'amas {\rouge du} gaz homogène sont {\rouge reliés à ceux dans le régulateur harmonique} par l'éq.~(\ref{eq:lienBb}).  Pour contrôler la troncature numérique dans la somme sur $\ell$ dans l'éq.~(\ref{eq:decomp_sur_l}), nous utilisons l'équivalent {\rouge à} grand $\ell$ résultant de la {\rouge note} 11 de \cite{CastinWernerCan} : 
\begin{equation}
\label{eq40}
\sigma_\ell \underset{\ell\to +\infty}{\sim} \frac{\eta (-1)^\ell}{\pi\sin 2\nu} Q_\ell\left(\frac{1}{\sin\nu}\right)
\end{equation} où $Q_\ell(z)=\int_{-1}^{1} \frac{\dd u}{2} \frac{P_\ell(u)}{z-u}$ est une fonction de Legendre de deuxième {\rouge espèce}\footnote{De la relation 8.723(2) de \cite{GR} on {\rouge tire} en outre {\rouge l'équivalent} $Q_\ell(1/\sin\nu)\sim(\pi/\ell)^{1/2} \tan^\ell(\nu/2)\sin(\nu/2)/\cos^{1/2}\nu$ pour $\ell\to+\infty$ mais garder {\rouge l'expression exacte de} $Q_\ell$ rend l'estimation {\rouge (\ref{eq40})} beaucoup plus précise à $\ell$ modérément élevé, pour {\rouge de petites valeurs de} $\nu$.}. Les {\rouge valeurs des} quantités pertinentes, telles que le rapport de masse, l'exposant d'Efimov $s_0$, l'échelle d'énergie globale $E_{\rm glob}$ et l'énergie du trimère de l'état fondamental $\epsilon_0$, sont {\rouge données} dans {\rouge la table}. La comparaison de $E_{\rm glob}$ et $-\epsilon_0 e^{2\pi/|s_0|}$ indique dans quelle mesure le trimère de l'état fondamental se trouve dans le régime de {\rouge portée nulle}, ce qui {\rouge devient} marginalement le cas pour le mélange rubidium-lithium. La connaissance de la valeur de $\epsilon_0$ donne également la température de {\rouge raccordement} $k_B T \approx 
|\epsilon_0|$ entre le régime dominé par le trimère à basse température $B_{2,1}\sim \exp(\beta |\epsilon_0|)$ et le régime fortement dissocié à haute température $B_{2,1}\approx \frac{|s_0|}{2\pi} \ln (\beta E_{\rm glob})$, voir \cite{CastinWernerCan} et notre {\rouge note} \ref{note:autre_forme}.

\section{Conclusion}

Nous avons considéré des mélanges binaires à l'équilibre thermique de particules bosoniques ou fermioniques sans interaction intra{\rouge -espèce}, mais avec {\rouge une} interaction inter-espèce unitaire, {\rouge c'est-à-dire} de longueur de diffusion infinie {\rouge dans l'onde} $s$ et de portée réelle et effective bien plus petite que la longueur d'onde de de Broglie thermique, une situation qui peut être réalisée expérimentalement avec des atomes froids. Les propriétés du système dépendent {\rouge alors} de manière cruciale du rapport de masse des deux espèces.

En généralisant {\rouge d}es résultats {\rouge connus}, nous avons obtenu, dans le cadre du modèle {\rouge de portée nulle}, des expressions analytiques pour les troisièmes coefficients d'{\rouge amas} ou du viriel, {\rouge en termes d'}intégrales du logarithme des fonctions transcendantes d'Efimov. Cela a été rendu possible par l'invariance d'échelle des conditions de contact de Wigner-Bethe-Peierls à deux corps. En général, le résultat dépend {\rouge de} paramètres à trois corps $R_t$ apparaissant dans les conditions de contact à trois corps, soit parce que l'effet Efimov {\rouge est présent} (un exposant d'échelle $s$ est imaginaire {\rouge pur}), soit parce que le système est dans le régime préefimovien (un exposant d'échelle $s$ est réel et proche de zéro, car le rapport de masse est proche d'un seuil de l'effet Efimov).

Pour prédire la valeur des paramètres à trois corps, nous avons {\rouge utilisé} le modèle microscopique d'une résonance de Feshbach infiniment étroite de longueur {\rouge de Feshbach} $R_*$, {\rouge ce} qui devrait également supprimer les pertes de particules à trois corps dans {\rouge les} expériences.  Nous avons ensuite obtenu une nouvelle expression analytique pour $R_t$, sous la forme d'une intégrale faisant intervenir à nouveau le logarithme de la fonction {\rouge transcendante} d'Efimov, à la fois dans les régimes efimovien et pr{\rouge é}efimovien. On constate que $R_t$ diverge pour $s=1/2$, {\rouge sans qu'il s'agisse} d'une véritable résonance à trois corps, car les conditions de contact à trois corps ne représentent fidèlement la véritable interaction microscopique qu'à des échelles d'énergie bien inférieures à $\hbar^2/m_r R_t^2$ (et à $\hbar^2/m_r R_*^2$, évidemment), avec $m_r$ la masse réduite des deux espèces. En particulier, dans le modèle de résonance de Feshbach étroite, il n'existe pas d'état trimère dans le régime pré{\rouge efim}ovien, contrairement à {\rouge ce que prédit le} modèle {\rouge de} portée nulle.

Enfin, nous avons appliqué ce travail analytique à {\rouge un} calcul explicite des {\rouge troisièmes} coefficients {\rouge d'}amas en fonction de la température pour toutes les combinaisons binaires de trois espèces atomiques bosoniques couramment utilisées dans les expériences {\rouge d'}atomes froids.

\section{Remerciements}
S. E. remercie le JSPS pour son soutien financier.

\end{document}